\documentclass[final,5p,times,twocolumn,authoryear]{elsarticle}
\usepackage{graphicx,times}             %for PS/EPS graphics inclusion, new
\usepackage{natbib}
\usepackage{amssymb,amsmath}
\usepackage{makecell}
\usepackage{float}
\usepackage{verbatim}
\usepackage{url}
\usepackage{multirow}
\usepackage{booktabs}
\usepackage{enumitem}
\usepackage{array}
\usepackage{tabularx} % 引入tabularx包

\usepackage{hyperref}
\hypersetup{colorlinks = true, linkcolor = green, anchorcolor = red, citecolor = blue, filecolor = red, pagecolor = red, urlcolor = red}

\journal{Astronomy $\&$ Computing}

\begin{document}

\begin{frontmatter}

%% Title, authors and addresses

%% use the tnoteref command within \title for footnotes;
%% use the tnotetext command for theassociated footnote;
%% use the fnref command within \author or \affiliation for footnotes;
%% use the fntext command for theassociated footnote;
%% use the corref command within \author for corresponding author footnotes;
%% use the cortext command for theassociated footnote;
%% use the ead command for the email address,
%% and the form \ead[url] for the home page:
%% \title{Title\tnoteref{label1}}
%% \tnotetext[label1]{}
%% \author{Name\corref{cor1}\fnref{label2}}
%% \ead{email address}
%% \ead[url]{home page}
%% \fntext[label2]{}
%% \cortext[cor1]{}
%% \affiliation{organization={},
%%            addressline={}, 
%%            city={},
%%            postcode={}, 
%%            state={},
%%            country={}}
%% \fntext[label3]{}

\title{Astronomical Knowledge Entity Extraction in Astrophysics Journal Articles via Large Language Models}

\author[1,2,3]{Wujun Shao}
\author[4]{Pengli Ji}
\author[1,2,6]{Dongwei Fan}
\author[5]{Yaohua Hu}
\author[4]{Xiaoran Yan}
\author[1,2,3]{Chenzhou Cui}
\author[1,2,3]{Linying Mi}
\author[1,2,3]{Lang Chen}
\author[4]{Rui Zhang\corref{zr}}
\cortext[zr]{Corresponding author}
\ead{ruizhang@zhejianglab.com}

\address[1]{National Astronomical Observatories, Chinese Academy of Sciences, Beijing 100012, People's Republic of China}
\address[2]{University of Chinese Academy of Sciences, Beijing 100049, People's Republic of China}
\address[3]{National Astronomical Data Center, Beijing 100012, People's Republic of China}
\address[4]{Research Institute of Artificial Intelligence, Zhejiang Lab, Hangzhou 311100, China People's Republic of China}
\address[5]{Xidian University, Xi’an 710126, China People's Republic of China}
\address[6]{Guilin University, Guangxi 541006, China People's Republic of China}

\begin{abstract}
%% Text of abstract
Astronomical knowledge entities, such as celestial object identifiers, are crucial for literature retrieval and knowledge graph construction, and other research and applications in the field of astronomy. Traditional methods of extracting knowledge entities from texts face challenges like high manual effort, poor generalization, and costly maintenance. Consequently, there is a pressing need for improved methods to efficiently extract them. This study explores the potential of pre-trained Large Language Models (LLMs) to perform astronomical knowledge entity extraction (KEE) task from astrophysical journal articles using prompts. We propose a prompting strategy called Prompt-KEE, which includes five prompt elements, and design eight combination prompts based on them. Celestial object identifier and telescope name, two most typical astronomical knowledge entities, are selected to be experimental object. And we introduce four currently representative LLMs, namely Llama-2-70B, GPT-3.5, GPT-4, and Claude 2. To accommodate their token limitations, we construct two datasets: the full texts and paragraph collections of 30 articles. Leveraging the eight prompts, we test on full texts with GPT-4 and Claude 2, on paragraph collections with all LLMs. The experimental results demonstrated that pre-trained LLMs have the significant potential to perform KEE tasks in astrophysics journal articles, but there are differences in their performance. Furthermore, we analyze some important factors that influence the performance of LLMs in entity extraction and provide insights for future KEE tasks in astrophysical articles using LLMs.
\end{abstract}

\begin{keyword}
%% keywords here, in the form: keyword \sep keyword, up to a maximum of 6 keywords
astronomical databases: miscellaneous

%% PACS codes here, in the form: \PACS code \sep code

%% MSC codes here, in the form: \MSC code \sep code
%% or \MSC[2008] code \sep code (2000 is the default)

\end{keyword}

\end{frontmatter}

%\tableofcontents

%% \linenumbers

%% main text

\section{Introduction}           %% first-level sections will be auto-capitalized
\label{sect:intro}

% In 1996, we started a project to obtain Johnson $V$ and Str\"{o}mgren
% $uvby\beta$ photometry for the poorly studied variables of
% ``pulsational interest''~$\cdots\cdots$.
% We used the three-channel high-speed photoelectric photometer designed for
% the Whole Earth Telescope campaign
% (\citealt{Nather+etal+1990, Jiang+Hu+1998}),
% and the four-channel Chevreton photoelectric photometer
% (\citealt{Michel+etal+1990, Michel+etal+1992}) dedicated to
% the STEPHI (STEllar Photometry International, \citealt{Michel+etal+1992}).

The advent of multi-band and multi-messenger observations marks a new epoch in the field of astronomy. This further attracted more and more scholars to devote themselves to astronomical research, resulting in an increasing cumulative amount of astronomical literature. Renowned repositories such as the Astrophysics Data System (ADS) \footnote{\url{https://ui.adsabs.harvard.edu/}} and ArXiv \footnote{\url{https://arxiv.org/archive/astro-ph/}} furnish a plethora of astrophysics journal articles, which are replete with invaluable specialized knowledge, including but not limited to, celestial object identifiers, and telescope names \citep{grezes2022overview}. These knowledge entities are crucial to the research and application of literature retrieval ~\citep{marrero2013named, yadav2019survey}, text mining, information association and recommendation, knowledge graph construction~\citep{hogan2021knowledge, al2020named}, publication management, etc. Effectively extracting these knowledge entities from the literature has become one of the keys to improve the efficiency and depth of astronomy research.

Knowledge Entity Extraction (KEE), a subtask of Named Entity Recognition (NER), emphasizes the extraction of professional knowledge entities from texts and outputs them in a structured format, negating the need for sequential entity annotation \citep{wang2023gpt}. Currently, KEE and NER are extensively studied across various domains. For instance, in the field of biology, this involves extracting information about genes, proteins, and biological processes from texts. In the medical field, it encompasses the identification of symptoms, diagnostic opinions, and drug information. In the field of astronomy, as astronomers' demand for entity information in texts increasingly grows, researchers have also embarked on some exploratory studies.

Early efforts in the field of astronomy have largely depended on rule-based methods \citep{cucerzan1999language, grishman1996message, cardie1997empirical} and dictionaries \citep{cohen2004exploiting, riloff1999learning, torisawa2007exploiting} for entity extraction. The DJIN (Journal of Identifiers and Names Detection) system is one example that uses the Dictionary of Nomenclature of Celestial Objects \citep{lortet1994second} to design more than 50,000 regular expressions for identifying celestial object identifiers and names in articles. This system's most successful implementation is evidenced at the Strasbourg Astronomical Data Centre (CDS), where it is seamlessly integrated with literature and catalog queries, providing a service within SIMBAD \footnote{\url{https://simbad.u-strasbg.fr/simbad/}} (Set of Identifications, Measurements, and Bibliography for Astronomical Data). This integration facilitates users in navigating through the celestial object identifiers and names mentioned in literature and accessing their detailed data during literature searches; conversely, it enables literature retrieval through celestial object identifiers and names. SIMBAD not only optimizes the process of knowledge acquisition for astronomers but also underscores the pivotal role of KEE in bridging various information resources within the field of astronomy.

With the development of machine learning \citep{mitchell1997machine, jordan2015machine, mahesh2020machine}, statistics-based NER, such as maximum entropy models \citep{bender2003maximum, curran2003language} and hidden Markov models \citep{shen2003effective, morwal2012named}, have emerged as the prevailing strategy. \cite{murphy2006named} pioneered the development of a specialized entity extraction system tailored for astronomical literature. Specifically, they utilized the maximum entropy model to train an application which is capable of identifying key entities such as source types, source names, and equipment names. By learning from extensive corpora, this approach can glean contextual and distributional information about entities, thereby enhancing recognition performance. However, these machine learning-based approaches are not without their limitations. They demonstrate a limited capacity to adapt to the increasingly diverse requirements of NER, coupled with a weak generalization ability. Moreover, they require substantial human intervention in aspects such as feature selection and entity annotation, leading to significant costs.

Employing Google's BERT (Bidirectional Encoder Representations from Transformers) ~\citep{devlin2018bert} deep neural network architecture, \cite{grezes2021building} have developed a domain-specific model for astronomy, termed astroBERT, through the training on a corpus comprising 395,499 astronomical research papers. Subsequently, this model was used for the development of the NER tool in ADS, which includes identifying specific organizations, projects, terms, etc., in the literature. Moreover, in the evaluation results, astroBERT performed better than the standard BERT model in the KEE tasks on ADS data. Following the success of this work, the Detection Entities in Astrophysics Literature (DEAL) shared task was proposed at the First Workshop on Information Extraction from Scientific Publications (WIESP)\footnote{\url{https://ui.adsabs.harvard.edu/WIESP/2022/SharedTasks}} at AACL-IJCNLP 2022 \citep{grezes2022overview}.
The DEAL challenge mandates participants to construct systems capable of automatically extracting astronomically named entities. Some researchers have attempted to use pre-trained language models such as mT5 \citep{ghosh2022astro} and BERT \citep{alkan2022majority} to extract knowledge entities from text, achieving considerable results.

Recently, LLMs with hundreds of billions of parameters, such as GPT-3.5, have demonstrated exceptional zero-shot and few-shot learning capabilities across a multitude of tasks \citep{wang2023gpt, li2023far, li2023prompt}. Owing to their extensive training samples, these models can rapidly comprehend the rich semantic knowledge embedded in text without the need for large annotated data. Their robust transfer learning capabilities also enable them to swiftly adapt to new domains. Therefore, large language models are also actively being applied to KEE tasks by researchers. \cite{sotnikov2023language} harnessed LLMs such as InstructGPT-3 ~\citep{ouyang2022training} and Flan-T5-XXL ~\citep{chung2022scaling} for the extraction of astronomical knowledge entities, including event IDs and object names, from Astronomical Telegrams and GCN Circulars. They explored various methods to enhance the capabilities of Large Language Models (LLMs), including prompt engineering and model fine-tuning. Their research highlights the potential of LLMs in NER tasks within the field of astronomy.

Owing to the increasing specialization and diversity of astronomical knowledge entities within articles, annotating and training copious samples for each type of entity to develop a functional extraction model is evidently inefficient and unsustainable. Therefore, in this paper, we focus on two representative knowledge entities within the field of astronomy: celestial object identifiers and astronomical telescopes. Furthermore, we select four mainstream LLMs (Llama-2-70B, GPT-3.5, GPT-4 and Claude 2) and carefully design a new strategy called Prompt-KEE to explore the potential of pre-trained LLMs for KEE in astrophysical articles.

The rest of this paper is structured as follows. In Section \ref{sec:method}, we describe the Prompt-KEE strategy and the four LLMs, as well as the set of prompts that we design based on this strategy. In Section \ref{sec:experiment}, we detail the dataset, the design of the combination prompts, the specific experimental procedures, and experimental results and analysis. In Section \ref{sec:discussion}, we briefly discuss our work. In Section \ref{sec:conclusion}, we conclude this paper.

\section{Method} \label{sec:method}
Astrophysical journal articles contain a wide variety of astronomical knowledge entities. Table ~\ref{tab:entity} shows two types of knowledge entities mentioned in sentences from different articles: celestial object identifiers and telescope names. Extracting them from articles is challenging. Inspired by NATURAL-INSTRUCTIONS \citep{mishra2021cross}, we propose a prompting strategy, Prompt-KEE, to explore the potential of using general LLMs to extract knowledge entities directly from astrophysical journal articles in a prompt-based method. In this section, we describe the five prompt components of Prompt-KEE, the specific prompts designed based on the Prompt-KEE framework, and four LLMs (Llama-2-70B, GPT-3.5, GPT-4 and Claude 2) used.

\begin{table*}[t]
    \caption{The example of sentences that contain celestial identifiers and telescope names}
    \label{tab:entity}
    \centering
    \renewcommand{\arraystretch}{2} % Increase the row height
    \begin{tabularx}{\textwidth}{p{0.15\linewidth} X p{0.2\linewidth}}
    \hline
    \textbf{Entity Category} & \textbf{Sentence} & \textbf{Reference} \\
    \hline
    \multirow{8}{*}{Object Identifier} & We performed a detailed chemical analysis for a few objects from this list and showed that the estimated abundances of the CEMP-r/s star \textbf{LAMOST J151003.74+305407.3} (hereafter \textbf{J151}) could be well explained by the model yields ([X/Fe]) of i-process nucleosynthesis of heavy elements, and \textbf{LAMOST J091608.81 + 230734.6} (hereafter \textbf{J091}) \ldots & \multirow{3}{*}{\citep{purandardas2022lamost}} \\
    \cline{2-3}
    & Only a handful of SySts exhibit noticeable signs of such variations in their SEDs (e.g., \textbf{2MASS J17391715-3546593}, \textbf{356.04+03.20}, \textbf{AS 245}, \textbf{H 2-34}, \textbf{PN H 2-5}, \textbf{RT Cru}, \textbf{SMP LMC 88}, \textbf{UV Aur}, \textbf{BI Cru}, \textbf{Hen 2-127}, \textbf{AS 221}, \textbf{Hen 2-139}, \textbf{K 3-9}, \textbf{RR Tel}, \textbf{V347 Nor}, \textbf{V835 Cen}, \textbf{354.98-02.87}). & \multirow{2.5}{*}{\citep{akras2019census}} \\
    \cline{2-3}
    & However, no apparent periods have been detected in the millisecond to second range for either \textbf{FRB 20121102A} or \textbf{FRB 20201124A}, two of the most well-studied repeaters \ldots & \multirow{2}{*}{\citep{niu2022fast}}\\
    \hline
    \multirow{6}{*}{Telescope Name} & \ldots which was identified from the \textbf{LAMOST} spectrum. The photometric data were collected with the \textbf{Tsinghua-NAOC 0.8 m telescope}(\textbf{TNT}), \textbf{Transiting Exoplanet Survey Satellite}(\textbf{TESS}), \textbf{Zwicky Transient Facility}(\textbf{ZTF}), and \textbf{ASAS-SN} \ldots & \multirow{2.5}{*}{\citep{li2023lamost}}\\
    \cline{2-3}
    & \textbf{Gaia} measurements of G29-38 will build on existing observations with \textbf{Keck}, the \textbf{Hubble Space Telescope}, \textbf{Herschel}, and \textbf{ALMA} \ldots & \multirow{1.5}{*}{\citep{sanderson2022can}} \\
    \cline{2-3}
    & The first observation for this pulsar was from the \textbf{Arecibo telescope} at 327 and 430 MHz\ldots, Although \textbf{FAST} is the largest and most sensitive radio telescope in the world \ldots & \multirow{2}{*}{\citep{shang2022bi}} \\
    \hline
    \end{tabularx}
\end{table*}

% \begin{figure*}[!tp]
% 	\centering 
% 	\includegraphics[width= 1 \textwidth]{entity.png}	
% 	\caption{The example of sentences that contain celestial identifiers and telescope names. These paragraphs, sequentially from top to bottom, are respectively referenced from \citep{purandardas2022lamost}, \citep{akras2019census}, \citep{niu2022fast}, \citep{li2023lamost}, \citep{sanderson2022can}, \citep{shang2022bi}.} 
% 	\label{fig:entity}
% \end{figure*}

\subsection{Prompt-KEE}

\begin{figure*}[!tp]
	\centering 
	\includegraphics[width= 0.9 \textwidth]{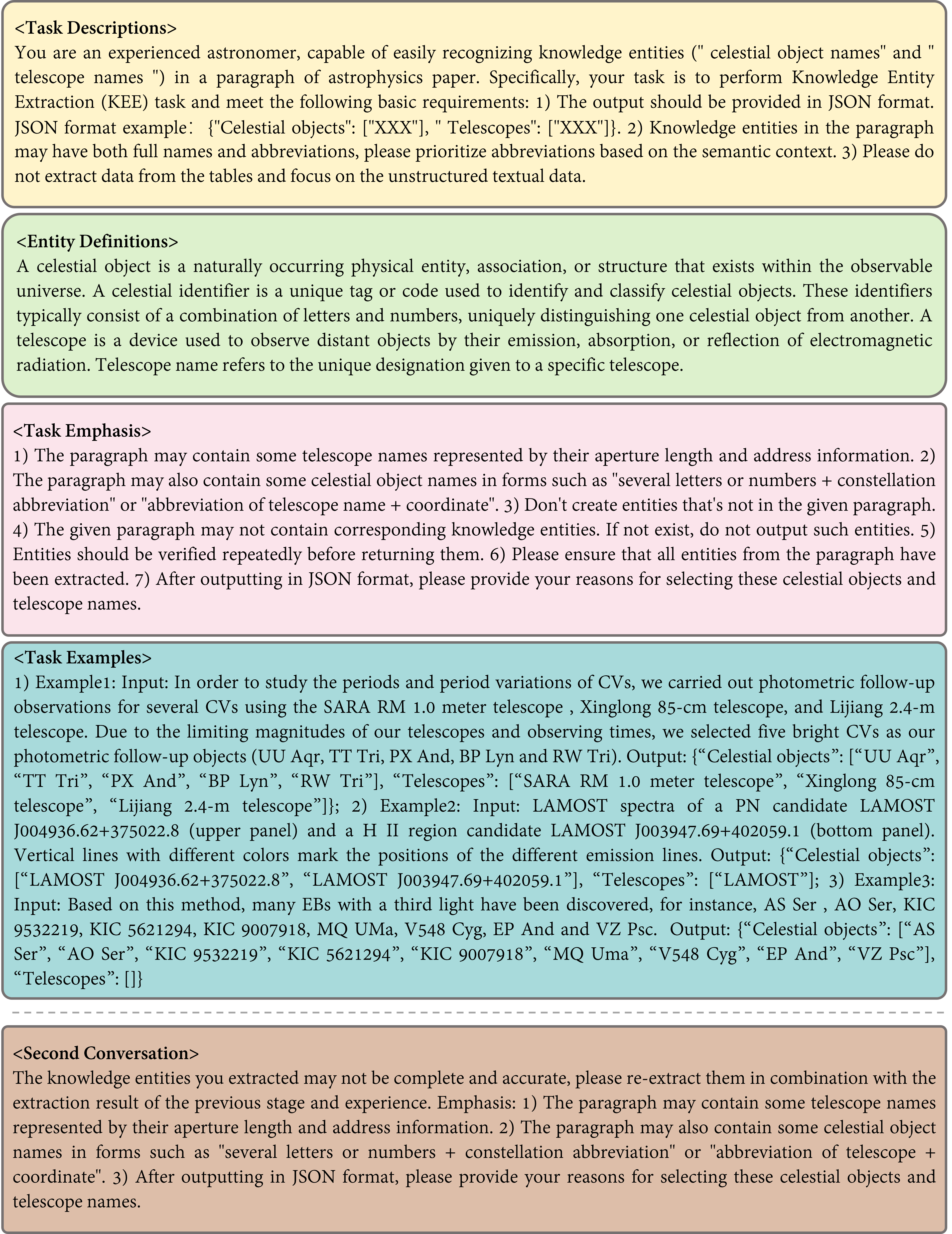}	
	\caption{A set of specific prompts that follow the Prompt-KEE strategy. The sentences from the three provided examples in the Task Examples are cited from \citep{han2018cataclysmic}, \citep{zhang2020catalogue}, and \citep{zhang20201swasp}.} 
	\label{fig:prompt}
\end{figure*}

Prompt-KEE is structured as a two-stage conversation process. In the first stage, the prompt comprises four components: Task Descriptions, Entity Definitions, Task Emphasis, and Task Examples. During the second stage, a partial utilization of Task Emphasis is employed specifically for the self-verification of LLMs. We follow the Prompt-KEE strategy to design a set of specific prompts, as shown in Figure \ref{fig:prompt}.

\subsubsection{Task Descriptions}
We carry out a general design of the Task Description to satisfy the subsequent comparative experiments. First, to fully exploit the astronomical knowledge captured by LLMs, we ask them to take on the role of an experienced astronomer and inform them about the working ability they need to master \citep{kong2023better}. Second, we explicitly specify that the task is to extract astronomical knowledge entities and output the results in a JSON format. Third, we provide three basic requirements. Considering that both the abbreviated and full forms of an astronomical knowledge entity might concurrently appear in an article, and professional scholars exhibit a preference for using abbreviations in academic writing, therefore, the first prompt asks LLMs to prioritize the extraction of entities in their abbreviated form. Most LLMs struggle with recognizing and processing structured tabular data \citep{bisercic2023interpretable}. Thus, our study skips any tables and figures, and give LLMs the other prompt, i.e., ignoring any information in tables or figures in the article.

\subsubsection{Entity Definitions}
LLMs often face challenges in distinguishing highly specialized and detailed terminology, as many similar terms can disperse LLMs' attention and thereby affect final performance \citep{zhao2023domain}. To address this issue, we provide definitions for the knowledge entities that we expect to be extracted (celestial object identifiers and telescope names) in an attempt to use these detailed definitions to guide LLMs in the distinction of professional terminology. Specifically, we reference the definitions of celestial object \footnote{\url{https://en.wikipedia.org/wiki/Astronomical_object}} and telescope \footnote{\url{https://en.wikipedia.org/wiki/Telescope}} from Wikipedia. We also define celestial object identifier and telescope name. Note that celestial objects may have different identifier formats (e.g., Vega, LAMOST J004936.62+375022.8, AS Ser) due to different naming conventions and standards, so we use celestial object identifier as a general term for the convenience of LLMs understanding. Additionally, we do not make a strict distinction between telescope names, other observational facility names, and sky survey names (e.g., LAMOST, Gaia, SDSS). In fact, astronomers usually do not emphasize the differences among them during scientific research.

\subsubsection{Task Emphasis}
Task Emphasis provides domain knowledge to LLMs for the target entities, and uses prompts to activate the self-improvement capabilities of LLMs. We indicate that telescopes may be named by their aperture length and address information, and celestial objects may be encoded as ``several letters or numbers + constellation abbreviation'' or ``abbreviation of telescope name + coordinate''. We employ self-check prompts to rectify potential errors in the output, thereby encouraging a more profound contextual understanding \citep{gero2023self}. It also contains some other prompts that should not be ignored, which come from our previous practical experience.

\subsubsection{Task Examples}
Task Examples are designed to enable LLMs to learn the mapping from the inputs to outputs for the celestial objects and telescope names KEE task \citep{min2022rethinking}. Task Examples are meticulously crafted to facilitate the learning process of LLMs in mapping inputs to outputs for celestial object identifiers and telescope names within the KEE task.
Specifically, the first example focuses on the form of telescope names, like ``address + aperture'' format often found in articles. The second one is designed for celestial object identifiers named in ``abbreviation of telescope + coordinate'' pattern. To reduce the likelihood of LLMs outputting non-existent entities, the third example uses an input without telescope names and an output with no telescope name as guidance.

\subsubsection{Second Conversation}
We assume that the extracted entities in the first stage may be incomplete and inaccurate. To make up for the errors and omissions of the extraction in the first stage, we provide another stage conversation prompts to ask LLMs to validate the results of the previous conversation and re-extract knowledge entities\citep{ji2023vicunaner}.

\subsection{LLMs Used}
Recent advancements in computational capabilities, combined with the accumulation of extensive textual datasets, have propelled the development of powerful LLMs. Notably, Llama-2-70B, GPT-3.5, GPT-4 and Claude 2 represent cutting-edge systems that have garnered significant attention.

Llama-2-70B \footnote{\url{https://ai.meta.com/llama/}}, a prominent member of the Llama 2 series \citep{touvron2023llama}, is characterized by its expansive 70 billion parameters and a 4096-token (100 tokens \~= 75 words) context window, making it adept at understanding and processing complex language structures. Its training on a massive 2 trillion token dataset significantly enhances its context comprehension, a critical factor in KEE tasks. The model's proficiency in external benchmarks, especially in areas requiring deep reasoning and knowledge understanding, positions it as an ideal candidate for testing KEE.

GPT-3.5 \footnote{\url{https://openai.com/chatgpt}}, an advanced natural language processing model from OpenAI built upon the GPT architecture, is a high-performance iteration that inherits the exceptional language generation and comprehension capabilities of GPT-3 \footnote{\url{https://openai.com/blog/gpt-3-apps}} while optimizing for faster response and lower cost through parameter efficiency. Through massive pre-training and self-supervised learning, this model has acquired mastery over linguistic patterns and structures, conferring immense application potential across diverse natural language processing tasks including text generation, question answering, knowledge entity extraction.

GPT-4, including its standard 8K and extended 32K token models, has been significantly advanced with the introduction of GPT-4 Turbo, capable of handling a 128K token context \footnote{\url{https://help.openai.com/en/articles/7127966-what-is-the-difference-between-the-gpt-4-models}}\footnote{\url{https://openai.com/blog/new-models-and-developer-products-announced-at-devday}}. This enhancement makes GPT-4 particularly adept at long-text KEE tasks. The extended token capacity allows for the processing of information equivalent to over 300 pages in a single prompt, thereby enabling more comprehensive analysis and understanding of extensive texts, vital for accurate and in-depth entity extraction from large documents or datasets.

Claude 2 \footnote{\url{https://claude.ai/chats}}, developed by Anthropic, is a LLM with a substantial token limit of 100,000 tokens. This extensive token capacity enables Claude 2 to handle and analyze large volumes of text in a single prompt, making it equally suitable for long-text KEE tasks. The ability to process such a high number of tokens allows Claude 2 to maintain context over lengthy documents, ensuring more accurate and comprehensive extraction of knowledge entities. This capability is crucial for analyzing and understanding extensive datasets or documents, where context and detailed comprehension are crucial.

\section{Experiments} \label{sec:experiment}
In this section, we first introduce the experimental datasets. Then, we describe the specific experimental settings. Finally, we present the evaluation metrics and experimental results.

\subsection{Dataset}

In order to evaluate the capability of LLMs in performing KEE tasks within astrophysics articles, we have established a dataset based on a set of specific selection criteria. Our focus was on selecting articles rich in distinct knowledge entities, such as celestial object identifiers and telescope names, to provide a diverse range of entity samples for the experiment. We ensured that the research subjects of the articles, covering galaxies, stars, planets, and more, as well as the observational bands like optical, radio, and X-ray, were as broad as possible. This breadth is crucial to fully reflect the diversity and complexity of astrophysical research. In addition, the selected articles needed to be logically structured and content-rich, which is essential for models to comprehend the text context and effectively identify and extract knowledge entities. By adhering to these criteria, our goal was to construct a dataset that truly showcases the potential of LLMs in conducting KEE tasks in astrophysical articles. 

Therefore, we carefully selected 30 astrophysics journal articles from authoritative publications, including the Astrophysical Journal (ApJ)\footnote{\url{https://iopscience.iop.org/journal/0004-637X}}, Astrophysical Journal Supplement Series \footnote{\url{https://iopscience.iop.org/journal/0067-0049}}, Astronomy \& Astrophysics (A\&A)\footnote{\url{https://www.aanda.org/}}, Monthly Notices of the Royal Astronomical Society (MNRAS)\footnote{\url{https://academic.oup.com/mnras/}}, and Research in Astronomy and Astrophysics (RAA)\footnote{\url{https://www.raa-journal.org/}}.

Despite GPT-4 and Claude 2's unparalleled long-context capabilities allowing them to effortlessly parse texts of the level of astrophysical journal articles, Llama-2-70B and GPT-3.5 currently face strict token limitations. Direct segmentation of lengthy texts based on the maximum token support of models can severely disrupt contextual semantic information, limiting the understanding and reasoning abilities of large language models. Considering that the maximum token supported by Llama-2-70B and GPT-3.5 is sufficient to cover each paragraph in the articles, we segmented all articles in the order of their paragraphs to ensure maximum preservation of contextual semantic integrity. After processing, these articles were divided into segments ranging from 20 to over 100 paragraphs, forming 30 paragraph collections.

In total, we collected two datasets: the full texts and paragraph collections of the 30 articles. Following the principle of prioritizing abbreviations, we annotated the celestial object identifiers and telescope names that appeared in them. The DOIs of the articles we selected and the specific annotation data are available in the Paperdata Repository of National Astronomical Data Center at \href{https://nadc.china-vo.org/res/r101358/}{https://nadc.china-vo.org/res/r101358/}.

\subsection{Experiment Setup} \label{subsec:comparative experiment}

\begin{figure*}[!tp]
	\centering 
	\includegraphics[width= 0.8 \textwidth]{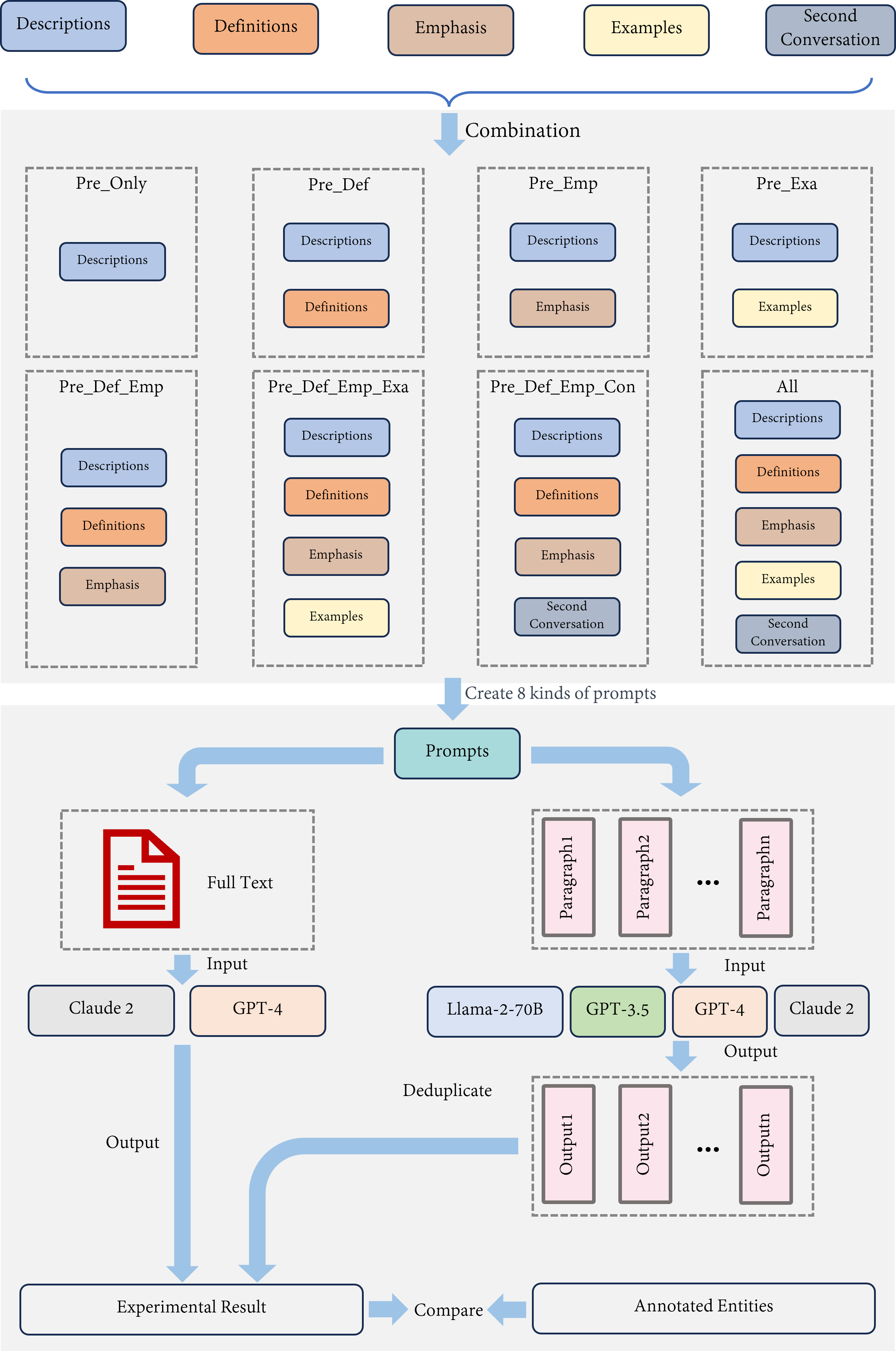}	
	\caption{The Experimental pipeline.} 
	\label{fig:step}
\end{figure*}

We designed the comparative experiment from the following aspects. And the Figure \ref{fig:step} illustrates our experimental pipeline. The terms Descriptions, Definitions, Emphasis, Examples, and Second Conversation in the figure represent Task Descriptions, Entity Definitions, Task Emphasis, Task Examples, and Second Conversation respectively in the prompt.

First, since we crafted Task Description in a general way, we combine Task Description with other prompt elements to explore the influence of each prompt element on KEE task. These combinations include: 1) Des\_Only: Task Descriptions only. 2) Des\_Def: Task Descriptions combined with Entity Definitions. 3) Des\_Emp: Task Descriptions combined with Task Emphasis. 4) Des\_Exa: Task Descriptions combined with Task Examples. 5) Des\_Def\_Emp: Task Descriptions, Entity Definitions, and Task Emphasis combined. 6) Des\_Def\_Emp\_Exa: Task Descriptions, Entity Definitions, Task Emphasis, and Task Examples combined. 7) Des\_Def\_Emp\_Con: Task Descriptions, Entity Definitions, Task Emphasis, and Second Conversation combined. 8) All: Task Descriptions, Entity Definitions, Task Emphasis, Task Examples, and Second Conversation combined. Based on our experience, we observed that the definitions of entities and the emphasis on the task usually have a positive and stable impact on the outputs of LLMs. However, the output performance of some current LLMs may exhibit uncertainty when examples are included \citep{zhao2021calibrate}. Therefore, we incorporated Task Examples in combinations 4), 6), and 8) to verify this possibility. Furthermore, Second Conversation is used to re-emphasize the task focus, leading us to accordingly construct combinations 7) and 8).

Second, we fed the full texts of 30 articles, each paired with the eight different combination prompts, into GPT-4 and Claude 2 for KEE. For the 30 paragraph collections of articles, we similarly combined them with these prompts and inputted them into Llama-2-70B, GPT-3.5, GPT-4, and Claude 2.  It's important to note that for Llama-2-70B and GPT-3.5, the knowledge entities extracted from the paragraph collections underwent a specific post-processing procedure, which involved merging the results and then removing duplicates.

Finally, all experimental results will be compared with the corresponding annotated knowledge entities of each article.

\subsection{Evaluation Metrics}

In the realm of KEE task, the evaluation metrics of Precision, Recall, and F1 Score are pivotal in ascertaining the efficacy of models in accurately discerning and extracting pertinent entities from textual data.

\textbf{Precision:} This metric quantifies the accuracy of the model in its entity extraction endeavors. Within the task of KEE, precision is defined as the proportion of accurately identified entities (TP) relative to the aggregate number of entities (TP+FP) extracted by the model. Elevated precision indicates a substantial ratio of correct entity identifications, signifying a reduction in false positives, i.e., erroneous extraction of non-entities or incorrect entities. The formula is
\begin{equation}
\textrm{Precision} = \frac{\textrm{TP}}{\textrm{TP}+\textrm{FP}}
\end{equation}

\textbf{Recall:} It assesses the model's ability to extract a comprehensive set of relevant entities. It measures the fraction of correctly extracted entities (TP) out of the total number of correct entities (TP+FN) that are inherently present and should be extracted from the textual corpus. A model exhibiting high recall is indicative of its proficiency in extracting the majority, if not the entirety, of pertinent entities, thereby minimizing instances of missed identifications. The formula is
\begin{equation}
\textrm{Recall} = \frac{\textrm{TP}}{\textrm{TP}+\textrm{FN}}
\end{equation}

\textbf{F1-score:} This metric, representing the harmonic mean of \textbf{Precision} and \textbf{Recall}, functions as an integrative measure that encapsulates both accuracy and completeness in KEE task. It is particularly salient in balancing the model's performance in avoiding entity omissions (high recall) and avoiding inaccurate extractions (high precision). The ideal scenario encompasses a model demonstrating concurrently high precision and recall, although a trade-off between these metrics is often observed in practical applications. The formula is
\begin{equation}
\textrm{F1-score} = 2 \cdot \frac{\textrm{Precision} \cdot \textrm{Recall}}{\textrm{Precision} + \textrm{Recall}}
\end{equation}

\subsection{Results and Analysis} \label{subsec:results and analysis}

In this section, we comprehensively evaluate the effectiveness of the Prompt-KEE strategy and the extraction capabilities of LLMs. Our analysis is bifurcated into two distinct parts on the full texts and paragraph collections.

\subsubsection{For Full Texts}

\begin{table*}[!th]
\caption{The results of GPT-4 and Claude 2 in extracting celestial object identifier and telescope name knowledge entities from the full text of 30 articles using each of the eight combination prompts individually
\label{tab:full_text}}
\begin{tabular*}{\textwidth}{l@{\extracolsep{\fill}}ccccccc}
\toprule
& \multirow{2}{*}{Combination Prompt} & \multicolumn{3}{c}{Celestial Object Identifier} & \multicolumn{3}{c}{Telescope Name} \\
\cline{3-5} \cline{6-8}
& & Precision & Recall & F1-score & Precision & Recall & F1-score \\
\midrule
\multirow{8}{*}{GPT-4}
& Des\_Only & 0.7913 & 0.4081 & 0.5385 & 0.8112 & 0.5179 & 0.6322 \\
& Des\_Def & 0.7813 & 0.4484 & 0.5698 & 0.8145 & 0.5449 & 0.6530 \\
& Des\_Emp & 0.8118 & 0.6480 & 0.7207 & 0.8540 & 0.7054 & 0.7726 \\
& Des\_Exa & 0.8309 & 0.5179 & 0.6381 & 0.8125 & 0.5223 & 0.6359 \\
& Des\_Def\_Emp & 0.8504 & 0.6502 & 0.7369 & 0.8549 & 0.7366 & 0.7914 \\
& Des\_Def\_Emp\_Exa & 0.8420 & 0.6569 & 0.7380 & 0.8684 & 0.7411 & 0.7997 \\
& Des\_Def\_Emp\_Con & 0.8713 & 0.6682 & 0.7563 & 0.8763 & 0.7589 & 0.8134 \\
& All & 0.8739 & 0.6839 & 0.7673 & 0.8769 & 0.7634 & 0.8162 \\
\midrule
\multirow{8}{*}{Claude 2}
& Des\_Only & 0.7456 & 0.1906 & 0.3036 & 0.6951 & 0.2545 & 0.3726 \\
& Des\_Def & 0.6912 & 0.2108 & 0.3231 & 0.7142 & 0.3571 & 0.4761 \\
& Des\_Emp & 0.7083 & 0.4193 & 0.5267 & 0.7952 & 0.5893 & 0.6769 \\
& Des\_Exa & 0.6987 & 0.3587 & 0.4703 & 0.7059 & 0.3750 & 0.4898 \\
& Des\_Def\_Emp & 0.7410 & 0.3722 & 0.4955 & 0.6927 & 0.6741 & 0.6833 \\
& Des\_Def\_Emp\_Exa & 0.7892 & 0.3946 & 0.5261 & 0.6748 & 0.7411 & 0.7064 \\
& Des\_Def\_Emp\_Con & 0.7500 & 0.4507 & 0.5630 & 0.7255 & 0.6964 & 0.7107 \\
& All & 0.7955 & 0.4798 & 0.5986 & 0.7652 & 0.7277 & 0.7459 \\
\bottomrule
\end{tabular*}
\end{table*}

\begin{figure}[t]
    \centering
    \includegraphics[width=0.48 \textwidth]{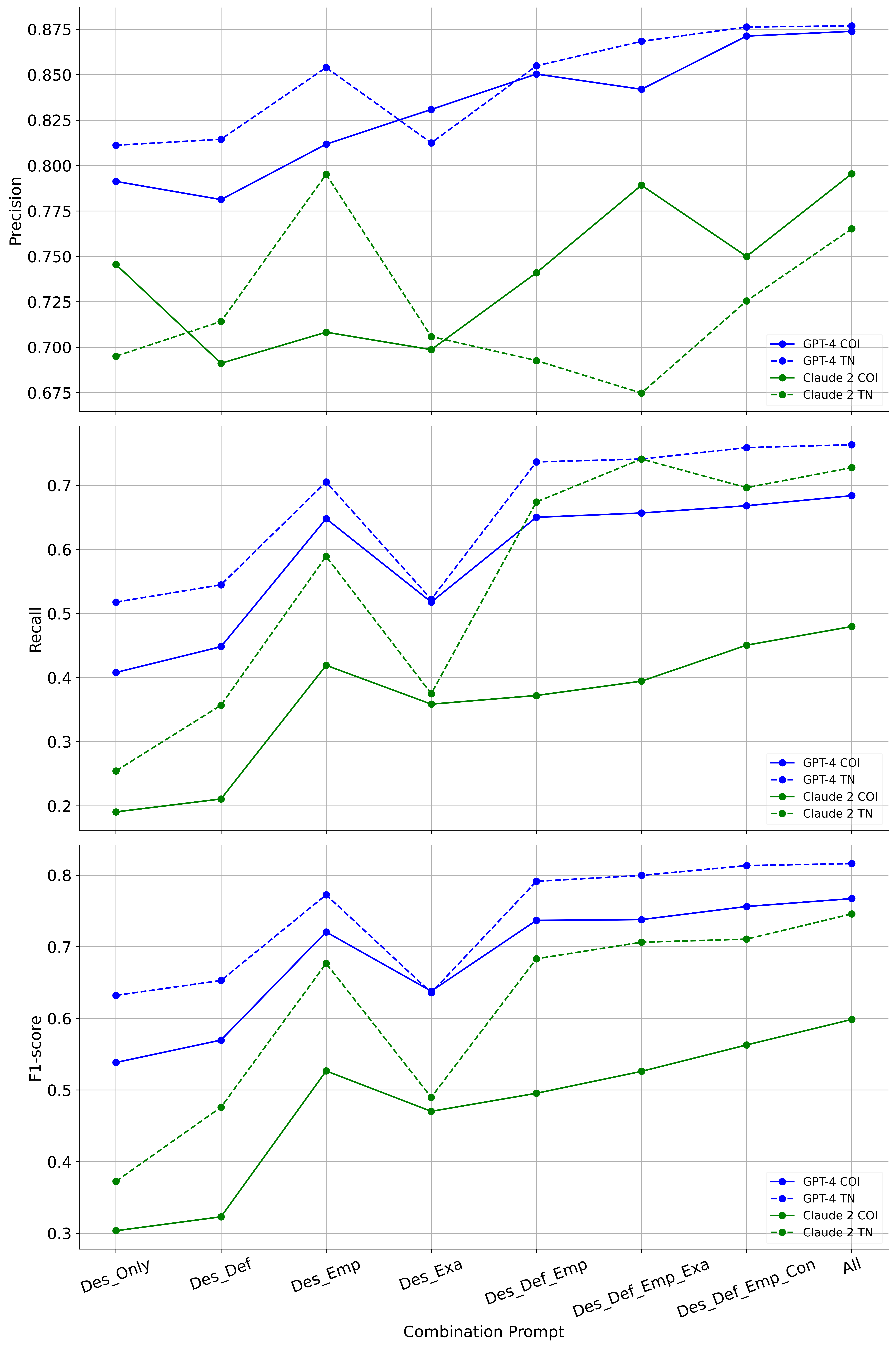}	
    \caption{The comparison of precision, recall, and F1-score for extracting celestial object identifiers and telescope names in the full texts between GPT-4 and Claude 2.}
    \label{fig:full_texts_results}
\end{figure}

As shown in Table~\ref{tab:full_text}, the results of GPT-4 and Claude 2 in extracting two types of knowledge entities (celestial object identifier and telescope name) from the full texts of the 30 articles are presented under eight different combination prompts. And in Figure \ref{fig:full_texts_results}, we compare precision, recall, and F1-score respectively. In a comparative analysis of these results, several key insights emerge.

GPT-4 consistently outperforms Claude 2 across all metrics. This superiority is particularly evident in the Recall and F1-scores, indicating GPT-4's great capability in accurately identifying a wider range of relevant entities. Claude 2, while competent, shows lower performance, especially in terms of Recall.

The diversity of prompts significantly influences the performance of both models. Task Descriptions integrating other elements, especially Task Emphasis, lead to better results compared to simpler ones. GPT-4 shows a pronounced ability to leverage diverse prompts for more accurate entity extraction, with the All prompt (encompassing a combination of all elements) yielding the highest F1-scores. Claude 2 also benefits from more elaborate prompts, but the improvement is less dramatic than with GPT-4. In addition, the overall improvement of the recall based on combination prompts is remarkably greater than the precision.

Moreover, a comparison between the extraction of celestial object identifiers and telescope names reveals GPT-4's relatively balanced performance for both entities, while Claude 2 exhibits more variance. GPT-4 slightly favors the extraction of telescope names, whereas Claude 2 shows a notable preference for telescope names, particularly in Recall and F1-score.

\subsubsection{For Paragraph Collections}

\begin{figure}[t!]
    \centering
    \includegraphics[width=0.48 \textwidth]{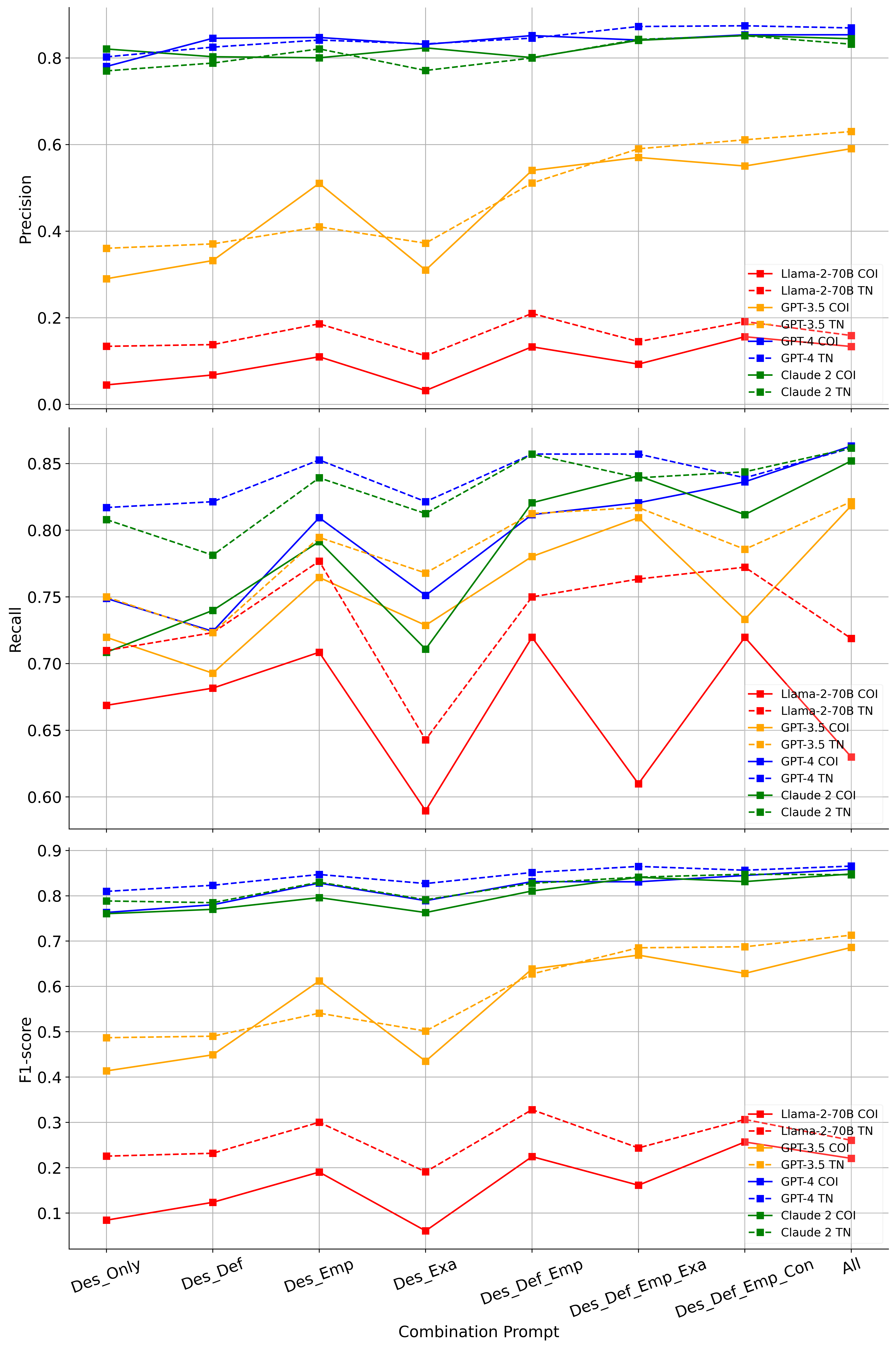}	
    \caption{The comparison of precision, recall, and F1-score for extracting celestial object identifiers and telescope names in the paragraph collections between Llama-2-70B, GPT-3.5, GPT-4, and Claude 2.}
    \label{fig:paragraph_results}
\end{figure}

The Table~\ref{tab:paragraph} provides the results of four models – Llama-2-70B, GPT-3.5, GPT-4, and Claude 2 – in extracting celestial object identifiers and telescope names from the paragraph collections of the articles. And in Figure \ref{fig:paragraph_results}, we compare three metrics respectively.

\begin{table*}[!th]
\caption{The results of Llama-2-70B, GPT-3.5, GPT-4 and Claude 2 in extracting celestial object identifier and telescope name knowledge entities from the paragraph collections of 30 articles using each of the eight combination prompts individually
\label{tab:paragraph}}
\begin{tabular*}{\textwidth}{l@{\extracolsep{\fill}}ccccccc}
\toprule
& \multirow{2}{*}{Combination Prompt} & \multicolumn{3}{c}{Celestial Object Identifier} & \multicolumn{3}{c}{Telescope Name} \\
\cline{3-5} \cline{6-8}
& & Precision & Recall & F1-score & Precision & Recall & F1-score \\
\midrule
\multirow{8}{*}{Llama-2-70B}
& Des\_Only & 0.0450 & 0.6687 & 0.0843 & 0.1341 & 0.7098 & 0.2256 \\
& Des\_Def & 0.0680 & 0.6816 & 0.1237 & 0.1381 & 0.7232 & 0.2319 \\
& Des\_Emp & 0.1100 & 0.7085 & 0.1904 & 0.1861 & 0.7768 & 0.3003 \\
& Des\_Exa & 0.0320 & 0.5897 & 0.0607 & 0.1121 & 0.6429 & 0.1909 \\
& Des\_Def\_Emp & 0.1330 & 0.7197 & 0.2245 & 0.2100 & 0.7500 & 0.3281 \\
& Des\_Def\_Emp\_Exa & 0.0930 & 0.6099 & 0.1614 & 0.1450 & 0.7634 & 0.2437 \\
& Des\_Def\_Emp\_Con & 0.1563 & 0.7197 & 0.2568 & 0.1912 & 0.7723 & 0.3065 \\
& All & 0.1337 & 0.6300 & 0.2206 & 0.1591 & 0.7188 & 0.2605 \\
\midrule
\multirow{8}{*}{GPT-3.5}
& Des\_Only & 0.2902 & 0.7197 & 0.4136 & 0.3605 & 0.7500 & 0.4869 \\
& Des\_Def & 0.3322 & 0.6928 & 0.4491 & 0.3707 & 0.7232 & 0.4902 \\
& Des\_Emp & 0.5105 & 0.7646 & 0.6122 & 0.4101 & 0.7946 & 0.5410 \\
& Des\_Exa & 0.3101 & 0.7287 & 0.4351 & 0.3723 & 0.7679 & 0.5015 \\
& Des\_Def\_Emp & 0.5404 & 0.7803 & 0.6386 & 0.5112 & 0.8125 & 0.6276 \\
& Des\_Def\_Emp\_Exa & 0.5703 & 0.8094 & 0.6691 & 0.5903 & 0.8170 & 0.6853 \\
& Des\_Def\_Emp\_Con & 0.5505 & 0.7332 & 0.6288 & 0.6111 & 0.7857 & 0.6875 \\
& All & 0.5906 & 0.8184 & 0.6861 & 0.6301 & 0.8214 & 0.7131 \\
\midrule
\multirow{8}{*}{GPT-4}
& Des\_Only & 0.7804 & 0.7489 & 0.7632 & 0.8026 & 0.8170 & 0.8097 \\
& Des\_Def & 0.8455 & 0.7242 & 0.7802 & 0.8251 & 0.8214 & 0.8232 \\
& Des\_Emp & 0.8474 & 0.8094 & 0.8280 & 0.8414 & 0.8527 & 0.8470 \\
& Des\_Exa & 0.8313 & 0.7511 & 0.7892 & 0.8326 & 0.8214 & 0.8270 \\
& Des\_Def\_Emp & 0.8518 & 0.8117 & 0.8313 & 0.8458 & 0.8571 & 0.8514 \\
& Des\_Def\_Emp\_Exa & 0.8414 & 0.8206 & 0.8309 & 0.8727 & 0.8571 & 0.8648 \\
& Des\_Def\_Emp\_Con & 0.8535 & 0.8363 & 0.8449 & 0.8744 & 0.8393 & 0.8564 \\
& All & 0.8536 & 0.8632 & 0.8584 & 0.8694 & 0.8616 & 0.8655 \\
\midrule
\multirow{8}{*}{Claude 2}
& Des\_Only & 0.8208 & 0.7085 & 0.7605 & 0.7702 & 0.8080 & 0.7886 \\
& Des\_Def & 0.8029 & 0.7399 & 0.7701 & 0.7883 & 0.7813 & 0.7848 \\
& Des\_Emp & 0.8005 & 0.7915 & 0.7960 & 0.8210 & 0.8393 & 0.8300 \\
& Des\_Exa & 0.8234 & 0.7108 & 0.7630 & 0.7712 & 0.8125 & 0.7913 \\
& Des\_Def\_Emp & 0.8009 & 0.8206 & 0.8106 & 0.8000 & 0.8571 & 0.8276 \\
& Des\_Def\_Emp\_Exa & 0.8408 & 0.8408 & 0.8408 & 0.8430 & 0.8393 & 0.8411 \\
& Des\_Def\_Emp\_Con & 0.8518 & 0.8117 & 0.8313 & 0.8514 & 0.8438 & 0.8475 \\
& All & 0.8444 & 0.8520 & 0.8482 & 0.8319 & 0.8616 & 0.8465 \\
\bottomrule
\end{tabular*}
\end{table*}

Llama-2-70B, while showing a respectable Recall, particularly in identifying telescope names, falls behind by a wide margin in precision, leading to lower F1-scores. This suggests Llama-2-70B's tendency to correctly identify a large number of relevant knowledge entities but at the cost of including more false positives. The pattern is consistent across all prompts, indicating a fundamental characteristic of the model's extraction strategy. Furthermore, the performance of Task Examples in Llama-2-70B validates the possibility that the example introduces uncertainty. This phenomenon arises from the fact that prompts with examples may bias the model's knowledge assessment, making it tend to prefer knowledge entities contained in the examples when extracting, while reduced the attention to information not present in the examples. 

GPT-3.5 demonstrates a marked improvement over Llama-2-70B, with notably higher Precision and F1-scores. Its performance peaks with the All prompt, suggesting an ability to effectively utilize prompt information for entity extraction. This trend implies that GPT-3.5 balances accuracy and comprehensiveness better than Llama-2-70B.

Claude 2, while slightly trailing behind GPT-4, shows impressive results. They outperforms both Llama-2-70B and GPT-3.5, exhibiting high scores in all metrics for both entity types. Their highest F1-scores are observed with the All prompt, indicating exceptional proficiency in handling complex prompts. And the results suggest an advanced understanding of the text and a more nuanced extraction capability, making them particularly suitable for tasks requiring high precision and recall.

By analyzing the F1-scores across above two tables, it is evident that the Prompt-KEE strategy significantly activates the ability of the four LLMs to identify celestial object identifiers and telescope names. Especially, the inclusion of Task Emphasis greatly improves the extracttion performance.

We also note that four LLMs are much better at recognizing telescope names than they are at identifying celestial object identifiers. This disparity can be attributed to the fact that telescope names often appear alongside distinctive vocabulary, such as ``telescope'', ``survey'', etc., which aids in the understanding and judgment of LLMs. Additionally, the number of telescope names is still within a manageable range, possibly already encompassed within the prior knowledge of LLMs. In contrast, celestial object identifiers are diverse in format and vast in quantity, posing a great challenge for LLMs.

\begin{figure}[t!]
    \centering
    \includegraphics[width=0.48 \textwidth]{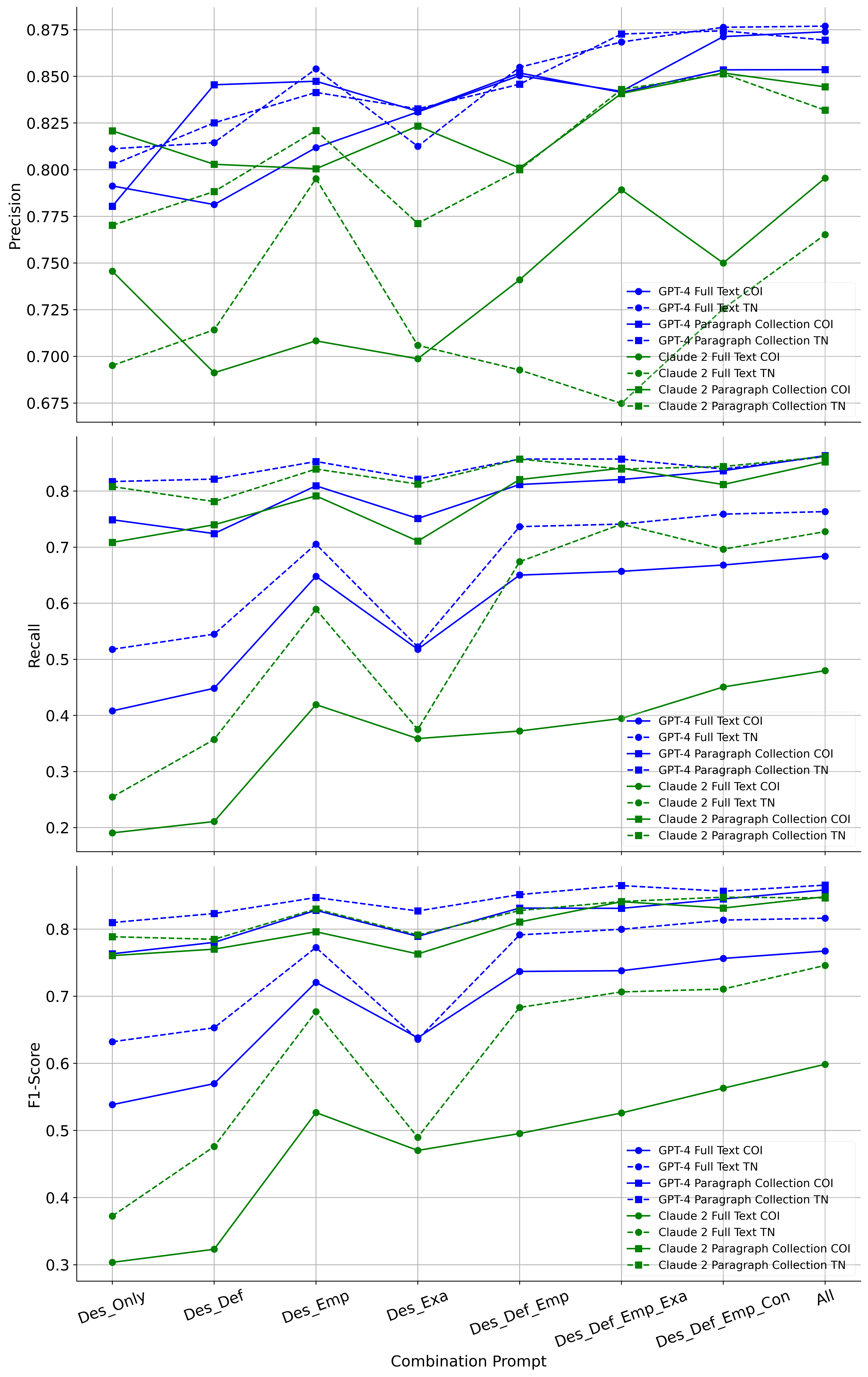}	
    \caption{The comparison of precision, recall, and F1-score for extracting celestial object identifiers and telescope names in the full texts and paragraph collections between GPT-4 and Claude 2.}
    \label{fig:two_results}
\end{figure}

Figure \ref{fig:two_results} compares the three evaluation metrics for extracting object identifiers and telescope names in the full texts and paragraph collections between GPT-4 and Claude 2. We observe distinct disparities in their performance across the texts of two different lengths. Specifically, the former consistently maintains high precision while showing the noticeable improvement in recall, whereas the latter demonstrates consistently high recall and precision. We attribute these differences primarily to the following factors:
\begin{enumerate}[label=\arabic*)]
    \item \textbf{Contextual Information}: The full texts of the 30 journal articles are imbued with a wealth of contextual information, which plays a pivotal role in enabling models like GPT-4 and Claude 2 to accurately comprehend the semantic nuances associated with celestial object identifiers and telescope names, thereby facilitating high precision in entity extraction. While paragraphs inherently provide a narrower context for knowledge entities, the sheer volume and diversity of training data underpinning GPT-4 and Claude 2 equip these models with the capability to maintain notable precision.
    \item \textbf{Entity Distribution}: Knowledge entities within full texts typically exhibit a sparser distribution pattern, whereas entities in paragraphs tend to be more concentrated. This distribution variance has some influence on the model's recognition capabilities. The sparse knowledge entities embedded within the full texts elevate the likelihood of models overlooking certain entities, consequently diminishing LLMs' recall. Conversely, paragraphs present a denser knowledge entity environment for the models. This attribute allows them to extract more potential knowledge entities, which greatly improves recall rates.
\end{enumerate}

These distinct attributes of full texts and paragraphs greatly affect how the models allocate their attention and process information, leading to varied performances in extracting knowledge entities from the texts of two different lengths within the astrophysical journal articles.

\section{Discussion} \label{sec:discussion}
This study validates the effectiveness of our carefully designed Prompt-KEE and highlights the potential of pre-trained LLMs for KEE in astrophysical journal articles. This further underscores the importance of prompts as a viable strategy to enable models to rapidly adapt to new domains and tasks. Although deploying these LLMs can be relatively complex or incur substantial API costs, it is very appealing as no additional annotations and training is needed. However, it is essential to acknowledge the limitations of this study. Firstly, we solely focused on two typical categories of astronomical knowledge entities, and the recognition of more complex and specialized astronomical entity types necessitates further investigation. Secondly, different models and astronomical task scenarios may require different prompt optimization strategies, and research in this area still offers extensive exploration opportunities.

Meanwhile, we also recognize several issues that warrant further investigation:
\begin{enumerate}[label=\arabic*)]
    \item Domain knowledge is key to augmenting the understanding, reasoning, and generalization abilities of LLMs; thus, training LLMs with extensive astronomical knowledge can further enhance their capability in extracting astronomical knowledge entities.
    \item The inclusion of examples may impact the model's extraction of new entities, suggesting that prompts need careful design based on specific models and scenarios. More information doesn't necessarily lead to better results, hence, further optimization of the Prompt-KEE strategy is possible.
    \item Differences in extraction results between full texts and paragraph collections, indicate that contextual information and dataset characteristics influence model attention allocation. Investigating how to guide the model to perform better across the texts of two different lengths merits further research.
    \item Despite the potential of LLMs for astronomical KEE, astronomy, as a specialized field, presents challenges for entity type and boundary recognition. This necessitates the development of domain-knowledge-enhanced strategies.
    \item Astrophysical journal articles often presents specialized knowledge entities such as celestial object identifiers in structured table. Existing models exhibit limited capabilities in extracting structured table data from articles. Future research can address this issue specifically.
\end{enumerate}

\section{Conclusion} \label{sec:conclusion}

In this paper, we proposed the Prompt-KEE strategy to explore the potential of pre-trained LLMs for KEE in astrophysical journal articles. We focused on the two most typical astronomical knowledge entities, celestial object identifier and telescope name. Based on the Prompt-KEE strategy, we designed eight combination prompts with five elements: Task Descriptions, Entity Definitions, Task Emphasis, Task Examples, and Second Conversation. Furthermore, we collected two datasets: the full texts and paragraph collections of the 30 articles, and employ four LLMs (Llama-2-70B, GPT-3.5, GPT-4 and Claude 2) for our experiments. Leveraging the eight combination prompts, we tested on full texts with GPT-4 and Claude-2, on paragraph collections with all LLMs. The experimental results demonstrated that pre-trained LLMs can perform KEE tasks in the astrophysics journal articles, but there are differences in their performance. Moreover, we introduced areas that require further exploration and improvement, including the design of prompts and the use of contextual information. This study provides valuable insights for using prompt engineering to adapt LLMs for KEE tasks in astrophysical articles.

% These experiments illustrate that Prompt-KEE strategy effectively activates the natural language understanding capability of large-scale language models in the field of astronomy.

\section{Acknowledgement}
This work is supported by the National Natural Science Foundation of China (NSFC)(12273077, 72101068, 12373110, 12103070), National Key Research and Development Program of China under Grants (2022YFF0712400, 2022YFF0711500), the 14th Five-year Informatization Plan of Chinese Academy of Sciences (CAS-WX2021SF-0204). Data resources are supported by China National Astronomical Data Center (NADC), CAS Astronomical Data Center and Chinese Virtual Observatory (China-VO). This work is supported by Astronomical Big Data Joint Research Center, co-founded by National Astronomical Observatories, Chinese Academy of Sciences and Alibaba Cloud.

%% If you have bibdatabase file and want bibtex to generate the
%% bibitems, please use
%%
\bibliographystyle{elsarticle-harv} 
\bibliography{example}

\begin{thebibliography}{48}
\expandafter\ifx\csname natexlab\endcsname\relax\def\natexlab#1{#1}\fi
\providecommand{\url}[1]{\texttt{#1}}
\providecommand{\href}[2]{#2}
\providecommand{\path}[1]{#1}
\providecommand{\DOIprefix}{doi:}
\providecommand{\ArXivprefix}{arXiv:}
\providecommand{\URLprefix}{URL: }
\providecommand{\Pubmedprefix}{pmid:}
\providecommand{\doi}[1]{\href{http://dx.doi.org/#1}{\path{#1}}}
\providecommand{\Pubmed}[1]{\href{pmid:#1}{\path{#1}}}
\providecommand{\bibinfo}[2]{#2}
\ifx\xfnm\relax \def\xfnm[#1]{\unskip,\space#1}\fi
%Type = Article
\bibitem[{Akras et~al.(2019)Akras, Guzman-Ramirez, Leal-Ferreira and
  Ramos-Larios}]{akras2019census}
\bibinfo{author}{Akras, S.}, \bibinfo{author}{Guzman-Ramirez, L.},
  \bibinfo{author}{Leal-Ferreira, M.L.}, \bibinfo{author}{Ramos-Larios, G.},
  \bibinfo{year}{2019}.
\newblock \bibinfo{title}{A census of symbiotic stars in the 2mass, wise, and
  gaia surveys}.
\newblock \bibinfo{journal}{The Astrophysical Journal Supplement Series}
  \bibinfo{volume}{240}, \bibinfo{pages}{21}.
%Type = Article
\bibitem[{Al-Moslmi et~al.(2020)Al-Moslmi, Oca{\~n}a, Opdahl and
  Veres}]{al2020named}
\bibinfo{author}{Al-Moslmi, T.}, \bibinfo{author}{Oca{\~n}a, M.G.},
  \bibinfo{author}{Opdahl, A.L.}, \bibinfo{author}{Veres, C.},
  \bibinfo{year}{2020}.
\newblock \bibinfo{title}{Named entity extraction for knowledge graphs: A
  literature overview}.
\newblock \bibinfo{journal}{IEEE Access} \bibinfo{volume}{8},
  \bibinfo{pages}{32862--32881}.
%Type = Inproceedings
\bibitem[{Alkan et~al.(2022)Alkan, Grouin, Schussler and
  Zweigenbaum}]{alkan2022majority}
\bibinfo{author}{Alkan, A.K.}, \bibinfo{author}{Grouin, C.},
  \bibinfo{author}{Schussler, F.}, \bibinfo{author}{Zweigenbaum, P.},
  \bibinfo{year}{2022}.
\newblock \bibinfo{title}{A majority voting strategy of a scibert-based
  ensemble models for detecting entities in the astrophysics literature (shared
  task)}, in: \bibinfo{booktitle}{Proceedings of the first Workshop on
  Information Extraction from Scientific Publications}, pp.
  \bibinfo{pages}{145--150}.
%Type = Inproceedings
\bibitem[{Bender et~al.(2003)Bender, Och and Ney}]{bender2003maximum}
\bibinfo{author}{Bender, O.}, \bibinfo{author}{Och, F.J.},
  \bibinfo{author}{Ney, H.}, \bibinfo{year}{2003}.
\newblock \bibinfo{title}{Maximum entropy models for named entity recognition},
  in: \bibinfo{booktitle}{Proceedings of the seventh conference on Natural
  language learning at HLT-NAACL 2003}, pp. \bibinfo{pages}{148--151}.
%Type = Article
\bibitem[{Bisercic et~al.(2023)Bisercic, Nikolic, van~der Schaar, Delibasic,
  Lio and Petrovic}]{bisercic2023interpretable}
\bibinfo{author}{Bisercic, A.}, \bibinfo{author}{Nikolic, M.},
  \bibinfo{author}{van~der Schaar, M.}, \bibinfo{author}{Delibasic, B.},
  \bibinfo{author}{Lio, P.}, \bibinfo{author}{Petrovic, A.},
  \bibinfo{year}{2023}.
\newblock \bibinfo{title}{Interpretable medical diagnostics with structured
  data extraction by large language models}.
\newblock \bibinfo{journal}{arXiv preprint arXiv:2306.05052} .
%Type = Article
\bibitem[{Cardie(1997)}]{cardie1997empirical}
\bibinfo{author}{Cardie, C.}, \bibinfo{year}{1997}.
\newblock \bibinfo{title}{Empirical methods in information extraction}.
\newblock \bibinfo{journal}{AI magazine} \bibinfo{volume}{18},
  \bibinfo{pages}{65--65}.
%Type = Article
\bibitem[{Chung et~al.(2022)Chung, Hou, Longpre, Zoph, Tay, Fedus, Li, Wang,
  Dehghani, Brahma et~al.}]{chung2022scaling}
\bibinfo{author}{Chung, H.W.}, \bibinfo{author}{Hou, L.},
  \bibinfo{author}{Longpre, S.}, \bibinfo{author}{Zoph, B.},
  \bibinfo{author}{Tay, Y.}, \bibinfo{author}{Fedus, W.}, \bibinfo{author}{Li,
  E.}, \bibinfo{author}{Wang, X.}, \bibinfo{author}{Dehghani, M.},
  \bibinfo{author}{Brahma, S.}, et~al., \bibinfo{year}{2022}.
\newblock \bibinfo{title}{Scaling instruction-finetuned language models}.
\newblock \bibinfo{journal}{arXiv preprint arXiv:2210.11416} .
%Type = Inproceedings
\bibitem[{Cohen and Sarawagi(2004)}]{cohen2004exploiting}
\bibinfo{author}{Cohen, W.W.}, \bibinfo{author}{Sarawagi, S.},
  \bibinfo{year}{2004}.
\newblock \bibinfo{title}{Exploiting dictionaries in named entity extraction:
  combining semi-markov extraction processes and data integration methods}, in:
  \bibinfo{booktitle}{Proceedings of the tenth ACM SIGKDD international
  conference on Knowledge discovery and data mining}, pp.
  \bibinfo{pages}{89--98}.
%Type = Inproceedings
\bibitem[{Cucerzan and Yarowsky(1999)}]{cucerzan1999language}
\bibinfo{author}{Cucerzan, S.}, \bibinfo{author}{Yarowsky, D.},
  \bibinfo{year}{1999}.
\newblock \bibinfo{title}{Language independent named entity recognition
  combining morphological and contextual evidence}, in:
  \bibinfo{booktitle}{1999 joint SIGDAT conference on empirical methods in
  natural language processing and very large corpora}.
%Type = Inproceedings
\bibitem[{Curran and Clark(2003)}]{curran2003language}
\bibinfo{author}{Curran, J.R.}, \bibinfo{author}{Clark, S.},
  \bibinfo{year}{2003}.
\newblock \bibinfo{title}{Language independent ner using a maximum entropy
  tagger}, in: \bibinfo{booktitle}{Proceedings of the seventh conference on
  Natural language learning at HLT-NAACL 2003}, pp. \bibinfo{pages}{164--167}.
%Type = Article
\bibitem[{Devlin et~al.(2018)Devlin, Chang, Lee and Toutanova}]{devlin2018bert}
\bibinfo{author}{Devlin, J.}, \bibinfo{author}{Chang, M.W.},
  \bibinfo{author}{Lee, K.}, \bibinfo{author}{Toutanova, K.},
  \bibinfo{year}{2018}.
\newblock \bibinfo{title}{Bert: Pre-training of deep bidirectional transformers
  for language understanding}.
\newblock \bibinfo{journal}{arXiv preprint arXiv:1810.04805} .
%Type = Article
\bibitem[{Gero et~al.(2023)Gero, Singh, Cheng, Naumann, Galley, Gao and
  Poon}]{gero2023self}
\bibinfo{author}{Gero, Z.}, \bibinfo{author}{Singh, C.},
  \bibinfo{author}{Cheng, H.}, \bibinfo{author}{Naumann, T.},
  \bibinfo{author}{Galley, M.}, \bibinfo{author}{Gao, J.},
  \bibinfo{author}{Poon, H.}, \bibinfo{year}{2023}.
\newblock \bibinfo{title}{Self-verification improves few-shot clinical
  information extraction}.
\newblock \bibinfo{journal}{arXiv preprint arXiv:2306.00024} .
%Type = Inproceedings
\bibitem[{Ghosh et~al.(2022)Ghosh, Santra, Iqbal and
  Basuchowdhuri}]{ghosh2022astro}
\bibinfo{author}{Ghosh, M.}, \bibinfo{author}{Santra, P.},
  \bibinfo{author}{Iqbal, S.A.}, \bibinfo{author}{Basuchowdhuri, P.},
  \bibinfo{year}{2022}.
\newblock \bibinfo{title}{Astro-mt5: Entity extraction from astrophysics
  literature using mt5 language model}, in: \bibinfo{booktitle}{Proceedings of
  the first Workshop on Information Extraction from Scientific Publications},
  pp. \bibinfo{pages}{100--104}.
%Type = Article
\bibitem[{Grezes et~al.(2021)Grezes, Blanco-Cuaresma, Accomazzi, Kurtz,
  Shapurian, Henneken, Grant, Thompson, Chyla, McDonald
  et~al.}]{grezes2021building}
\bibinfo{author}{Grezes, F.}, \bibinfo{author}{Blanco-Cuaresma, S.},
  \bibinfo{author}{Accomazzi, A.}, \bibinfo{author}{Kurtz, M.J.},
  \bibinfo{author}{Shapurian, G.}, \bibinfo{author}{Henneken, E.},
  \bibinfo{author}{Grant, C.S.}, \bibinfo{author}{Thompson, D.M.},
  \bibinfo{author}{Chyla, R.}, \bibinfo{author}{McDonald, S.}, et~al.,
  \bibinfo{year}{2021}.
\newblock \bibinfo{title}{Building astrobert, a language model for astronomy \&
  astrophysics}.
\newblock \bibinfo{journal}{arXiv preprint arXiv:2112.00590} .
%Type = Inproceedings
\bibitem[{Grezes et~al.(2022)Grezes, Blanco-Cuaresma, Allen and
  Ghosal}]{grezes2022overview}
\bibinfo{author}{Grezes, F.}, \bibinfo{author}{Blanco-Cuaresma, S.},
  \bibinfo{author}{Allen, T.}, \bibinfo{author}{Ghosal, T.},
  \bibinfo{year}{2022}.
\newblock \bibinfo{title}{Overview of the first shared task on detecting
  entities in the astrophysics literature (deal)}, in:
  \bibinfo{booktitle}{Proceedings of the first Workshop on Information
  Extraction from Scientific Publications}, pp. \bibinfo{pages}{1--7}.
%Type = Inproceedings
\bibitem[{Grishman and Sundheim(1996)}]{grishman1996message}
\bibinfo{author}{Grishman, R.}, \bibinfo{author}{Sundheim, B.M.},
  \bibinfo{year}{1996}.
\newblock \bibinfo{title}{Message understanding conference-6: A brief history},
  in: \bibinfo{booktitle}{COLING 1996 Volume 1: The 16th International
  Conference on Computational Linguistics}.
%Type = Article
\bibitem[{Han et~al.(2018)Han, Zhang, Shi, Pi, Lu, Zhao, Terheide and
  Jiang}]{han2018cataclysmic}
\bibinfo{author}{Han, X.L.}, \bibinfo{author}{Zhang, L.Y.},
  \bibinfo{author}{Shi, J.R.}, \bibinfo{author}{Pi, Q.F.}, \bibinfo{author}{Lu,
  H.P.}, \bibinfo{author}{Zhao, L.B.}, \bibinfo{author}{Terheide, R.K.},
  \bibinfo{author}{Jiang, L.Y.}, \bibinfo{year}{2018}.
\newblock \bibinfo{title}{Cataclysmic variables based on the stellar spectral
  survey lamost dr3}.
\newblock \bibinfo{journal}{Research in Astronomy and Astrophysics}
  \bibinfo{volume}{18}, \bibinfo{pages}{068}.
%Type = Article
\bibitem[{Hogan et~al.(2021)Hogan, Blomqvist, Cochez, d’Amato, Melo,
  Gutierrez, Kirrane, Gayo, Navigli, Neumaier et~al.}]{hogan2021knowledge}
\bibinfo{author}{Hogan, A.}, \bibinfo{author}{Blomqvist, E.},
  \bibinfo{author}{Cochez, M.}, \bibinfo{author}{d’Amato, C.},
  \bibinfo{author}{Melo, G.D.}, \bibinfo{author}{Gutierrez, C.},
  \bibinfo{author}{Kirrane, S.}, \bibinfo{author}{Gayo, J.E.L.},
  \bibinfo{author}{Navigli, R.}, \bibinfo{author}{Neumaier, S.}, et~al.,
  \bibinfo{year}{2021}.
\newblock \bibinfo{title}{Knowledge graphs}.
\newblock \bibinfo{journal}{ACM Computing Surveys (Csur)} \bibinfo{volume}{54},
  \bibinfo{pages}{1--37}.
%Type = Article
\bibitem[{Ji(2023)}]{ji2023vicunaner}
\bibinfo{author}{Ji, B.}, \bibinfo{year}{2023}.
\newblock \bibinfo{title}{Vicunaner: Zero/few-shot named entity recognition
  using vicuna}.
\newblock \bibinfo{journal}{arXiv preprint arXiv:2305.03253} .
%Type = Article
\bibitem[{Jordan and Mitchell(2015)}]{jordan2015machine}
\bibinfo{author}{Jordan, M.I.}, \bibinfo{author}{Mitchell, T.M.},
  \bibinfo{year}{2015}.
\newblock \bibinfo{title}{Machine learning: Trends, perspectives, and
  prospects}.
\newblock \bibinfo{journal}{Science} \bibinfo{volume}{349},
  \bibinfo{pages}{255--260}.
%Type = Article
\bibitem[{Kong et~al.(2023)Kong, Zhao, Chen, Li, Qin, Sun and
  Zhou}]{kong2023better}
\bibinfo{author}{Kong, A.}, \bibinfo{author}{Zhao, S.}, \bibinfo{author}{Chen,
  H.}, \bibinfo{author}{Li, Q.}, \bibinfo{author}{Qin, Y.},
  \bibinfo{author}{Sun, R.}, \bibinfo{author}{Zhou, X.}, \bibinfo{year}{2023}.
\newblock \bibinfo{title}{Better zero-shot reasoning with role-play prompting}.
\newblock \bibinfo{journal}{arXiv preprint arXiv:2308.07702} .
%Type = Article
\bibitem[{Li et~al.(2023a)Li, Li, Pan and Pan}]{li2023prompt}
\bibinfo{author}{Li, J.}, \bibinfo{author}{Li, H.}, \bibinfo{author}{Pan, Z.},
  \bibinfo{author}{Pan, G.}, \bibinfo{year}{2023}a.
\newblock \bibinfo{title}{Prompt chatgpt in mner: Improved multimodal named
  entity recognition method based on auxiliary refining knowledge from
  chatgpt}.
\newblock \bibinfo{journal}{arXiv preprint arXiv:2305.12212} .
%Type = Article
\bibitem[{Li and Zhang(2023)}]{li2023far}
\bibinfo{author}{Li, M.}, \bibinfo{author}{Zhang, R.}, \bibinfo{year}{2023}.
\newblock \bibinfo{title}{How far is language model from 100\% few-shot named
  entity recognition in medical domain}.
\newblock \bibinfo{journal}{arXiv preprint arXiv:2307.00186} .
%Type = Article
\bibitem[{Li et~al.(2023b)Li, Wang, Liu, Guo, Zhang, Sun, Song and
  Liu}]{li2023lamost}
\bibinfo{author}{Li, X.}, \bibinfo{author}{Wang, X.}, \bibinfo{author}{Liu,
  J.}, \bibinfo{author}{Guo, J.}, \bibinfo{author}{Zhang, Z.},
  \bibinfo{author}{Sun, Y.}, \bibinfo{author}{Song, X.}, \bibinfo{author}{Liu,
  C.}, \bibinfo{year}{2023}b.
\newblock \bibinfo{title}{Lamost j2043+ 3413--a fast disk precession sw sextans
  candidate in period gap}.
\newblock \bibinfo{journal}{arXiv preprint arXiv:2306.07529} .
%Type = Article
\bibitem[{Lortet et~al.(1994)Lortet, Borde and Ochsenbein}]{lortet1994second}
\bibinfo{author}{Lortet, M.C.}, \bibinfo{author}{Borde, S.},
  \bibinfo{author}{Ochsenbein, F.}, \bibinfo{year}{1994}.
\newblock \bibinfo{title}{Second reference dictionary of the nomenclature of
  celestial objects.}
\newblock \bibinfo{journal}{Astronomy and Astrophysics Suppl., Vol. 107, p.
  193-218 (1994)} \bibinfo{volume}{107}, \bibinfo{pages}{193--218}.
%Type = Article
\bibitem[{Mahesh(2020)}]{mahesh2020machine}
\bibinfo{author}{Mahesh, B.}, \bibinfo{year}{2020}.
\newblock \bibinfo{title}{Machine learning algorithms-a review}.
\newblock \bibinfo{journal}{International Journal of Science and Research
  (IJSR).[Internet]} \bibinfo{volume}{9}, \bibinfo{pages}{381--386}.
%Type = Article
\bibitem[{Marrero et~al.(2013)Marrero, Urbano, S{\'a}nchez-Cuadrado, Morato and
  G{\'o}mez-Berb{\'\i}s}]{marrero2013named}
\bibinfo{author}{Marrero, M.}, \bibinfo{author}{Urbano, J.},
  \bibinfo{author}{S{\'a}nchez-Cuadrado, S.}, \bibinfo{author}{Morato, J.},
  \bibinfo{author}{G{\'o}mez-Berb{\'\i}s, J.M.}, \bibinfo{year}{2013}.
\newblock \bibinfo{title}{Named entity recognition: fallacies, challenges and
  opportunities}.
\newblock \bibinfo{journal}{Computer Standards \& Interfaces}
  \bibinfo{volume}{35}, \bibinfo{pages}{482--489}.
%Type = Article
\bibitem[{Min et~al.(2022)Min, Lyu, Holtzman, Artetxe, Lewis, Hajishirzi and
  Zettlemoyer}]{min2022rethinking}
\bibinfo{author}{Min, S.}, \bibinfo{author}{Lyu, X.},
  \bibinfo{author}{Holtzman, A.}, \bibinfo{author}{Artetxe, M.},
  \bibinfo{author}{Lewis, M.}, \bibinfo{author}{Hajishirzi, H.},
  \bibinfo{author}{Zettlemoyer, L.}, \bibinfo{year}{2022}.
\newblock \bibinfo{title}{Rethinking the role of demonstrations: What makes
  in-context learning work?}
\newblock \bibinfo{journal}{arXiv preprint arXiv:2202.12837} .
%Type = Article
\bibitem[{Mishra et~al.(2021)Mishra, Khashabi, Baral and
  Hajishirzi}]{mishra2021cross}
\bibinfo{author}{Mishra, S.}, \bibinfo{author}{Khashabi, D.},
  \bibinfo{author}{Baral, C.}, \bibinfo{author}{Hajishirzi, H.},
  \bibinfo{year}{2021}.
\newblock \bibinfo{title}{Cross-task generalization via natural language
  crowdsourcing instructions}.
\newblock \bibinfo{journal}{arXiv preprint arXiv:2104.08773} .
%Type = Misc
\bibitem[{Mitchell(1997)}]{mitchell1997machine}
\bibinfo{author}{Mitchell, T.M.}, \bibinfo{year}{1997}.
\newblock \bibinfo{title}{Machine learning}.
%Type = Article
\bibitem[{Morwal et~al.(2012)Morwal, Jahan and Chopra}]{morwal2012named}
\bibinfo{author}{Morwal, S.}, \bibinfo{author}{Jahan, N.},
  \bibinfo{author}{Chopra, D.}, \bibinfo{year}{2012}.
\newblock \bibinfo{title}{Named entity recognition using hidden markov model
  (hmm)}.
\newblock \bibinfo{journal}{International Journal on Natural Language Computing
  (IJNLC) Vol} \bibinfo{volume}{1}.
%Type = Inproceedings
\bibitem[{Murphy et~al.(2006)Murphy, McIntosh and Curran}]{murphy2006named}
\bibinfo{author}{Murphy, T.}, \bibinfo{author}{McIntosh, T.},
  \bibinfo{author}{Curran, J.R.}, \bibinfo{year}{2006}.
\newblock \bibinfo{title}{Named entity recognition for astronomy literature},
  in: \bibinfo{booktitle}{Proceedings of the Australasian Language Technology
  Workshop 2006}, pp. \bibinfo{pages}{59--66}.
%Type = Article
\bibitem[{Niu et~al.(2022)Niu, Zhu, Zhang, Yuan, Zhou, Zhang, Jiang, Han, Li,
  Lee et~al.}]{niu2022fast}
\bibinfo{author}{Niu, J.R.}, \bibinfo{author}{Zhu, W.W.},
  \bibinfo{author}{Zhang, B.}, \bibinfo{author}{Yuan, M.},
  \bibinfo{author}{Zhou, D.J.}, \bibinfo{author}{Zhang, Y.K.},
  \bibinfo{author}{Jiang, J.C.}, \bibinfo{author}{Han, J.},
  \bibinfo{author}{Li, D.}, \bibinfo{author}{Lee, K.J.}, et~al.,
  \bibinfo{year}{2022}.
\newblock \bibinfo{title}{Fast observations of an extremely active episode of
  frb 20201124a. iv. spin period search}.
\newblock \bibinfo{journal}{Research in Astronomy and Astrophysics}
  \bibinfo{volume}{22}, \bibinfo{pages}{124004}.
%Type = Article
\bibitem[{Ouyang et~al.(2022)Ouyang, Wu, Jiang, Almeida, Wainwright, Mishkin,
  Zhang, Agarwal, Slama, Ray et~al.}]{ouyang2022training}
\bibinfo{author}{Ouyang, L.}, \bibinfo{author}{Wu, J.}, \bibinfo{author}{Jiang,
  X.}, \bibinfo{author}{Almeida, D.}, \bibinfo{author}{Wainwright, C.},
  \bibinfo{author}{Mishkin, P.}, \bibinfo{author}{Zhang, C.},
  \bibinfo{author}{Agarwal, S.}, \bibinfo{author}{Slama, K.},
  \bibinfo{author}{Ray, A.}, et~al., \bibinfo{year}{2022}.
\newblock \bibinfo{title}{Training language models to follow instructions with
  human feedback}.
\newblock \bibinfo{journal}{Advances in Neural Information Processing Systems}
  \bibinfo{volume}{35}, \bibinfo{pages}{27730--27744}.
%Type = Article
\bibitem[{Purandardas et~al.(2022)Purandardas, Goswami, Shejeelammal, Sonamben,
  Pawar, Mkrtichian, Doddamani and Joshi}]{purandardas2022lamost}
\bibinfo{author}{Purandardas, M.}, \bibinfo{author}{Goswami, A.},
  \bibinfo{author}{Shejeelammal, J.}, \bibinfo{author}{Sonamben, M.},
  \bibinfo{author}{Pawar, G.}, \bibinfo{author}{Mkrtichian, D.},
  \bibinfo{author}{Doddamani, V.H.}, \bibinfo{author}{Joshi, S.},
  \bibinfo{year}{2022}.
\newblock \bibinfo{title}{Lamost j045019. 27+ 394758.7, with peculiar
  abundances of n, na, v, zn, is possibly a sculptor dwarf galaxy escapee}.
\newblock \bibinfo{journal}{Monthly Notices of the Royal Astronomical Society}
  \bibinfo{volume}{513}, \bibinfo{pages}{4696--4710}.
%Type = Inproceedings
\bibitem[{Riloff et~al.(1999)Riloff, Jones et~al.}]{riloff1999learning}
\bibinfo{author}{Riloff, E.}, \bibinfo{author}{Jones, R.}, et~al.,
  \bibinfo{year}{1999}.
\newblock \bibinfo{title}{Learning dictionaries for information extraction by
  multi-level bootstrapping}, in: \bibinfo{booktitle}{AAAI/IAAI}, pp.
  \bibinfo{pages}{474--479}.
%Type = Article
\bibitem[{Sanderson et~al.(2022)Sanderson, Bonsor and
  Mustill}]{sanderson2022can}
\bibinfo{author}{Sanderson, H.}, \bibinfo{author}{Bonsor, A.},
  \bibinfo{author}{Mustill, A.}, \bibinfo{year}{2022}.
\newblock \bibinfo{title}{Can gaia find planets around white dwarfs?}
\newblock \bibinfo{journal}{Monthly Notices of the Royal Astronomical Society}
  \bibinfo{volume}{517}, \bibinfo{pages}{5835--5852}.
%Type = Article
\bibitem[{Shang et~al.(2022)Shang, Bai, Dang and Zhi}]{shang2022bi}
\bibinfo{author}{Shang, L.H.}, \bibinfo{author}{Bai, J.T.},
  \bibinfo{author}{Dang, S.J.}, \bibinfo{author}{Zhi, Q.J.},
  \bibinfo{year}{2022}.
\newblock \bibinfo{title}{The “bi-drifting” subpulses of psr j0815+ 0939
  observed with the five-hundred-meter aperture spherical radio telescope}.
\newblock \bibinfo{journal}{Research in Astronomy and Astrophysics}
  \bibinfo{volume}{22}, \bibinfo{pages}{025018}.
%Type = Inproceedings
\bibitem[{Shen et~al.(2003)Shen, Zhang, Zhou, Su and Tan}]{shen2003effective}
\bibinfo{author}{Shen, D.}, \bibinfo{author}{Zhang, J.}, \bibinfo{author}{Zhou,
  G.}, \bibinfo{author}{Su, J.}, \bibinfo{author}{Tan, C.L.},
  \bibinfo{year}{2003}.
\newblock \bibinfo{title}{Effective adaptation of hidden markov model-based
  named entity recognizer for biomedical domain}, in:
  \bibinfo{booktitle}{Proceedings of the ACL 2003 workshop on Natural language
  processing in biomedicine}, pp. \bibinfo{pages}{49--56}.
%Type = Article
\bibitem[{Sotnikov and Chaikova(2023)}]{sotnikov2023language}
\bibinfo{author}{Sotnikov, V.}, \bibinfo{author}{Chaikova, A.},
  \bibinfo{year}{2023}.
\newblock \bibinfo{title}{Language models for multimessenger astronomy}.
\newblock \bibinfo{journal}{Galaxies} \bibinfo{volume}{11},
  \bibinfo{pages}{63}.
%Type = Inproceedings
\bibitem[{Torisawa et~al.(2007)}]{torisawa2007exploiting}
\bibinfo{author}{Torisawa, K.}, et~al., \bibinfo{year}{2007}.
\newblock \bibinfo{title}{Exploiting wikipedia as external knowledge for named
  entity recognition}, in: \bibinfo{booktitle}{Proceedings of the 2007 joint
  conference on empirical methods in natural language processing and
  computational natural language learning (EMNLP-CoNLL)}, pp.
  \bibinfo{pages}{698--707}.
%Type = Article
\bibitem[{Touvron et~al.(2023)Touvron, Martin, Stone, Albert, Almahairi,
  Babaei, Bashlykov, Batra, Bhargava, Bhosale et~al.}]{touvron2023llama}
\bibinfo{author}{Touvron, H.}, \bibinfo{author}{Martin, L.},
  \bibinfo{author}{Stone, K.}, \bibinfo{author}{Albert, P.},
  \bibinfo{author}{Almahairi, A.}, \bibinfo{author}{Babaei, Y.},
  \bibinfo{author}{Bashlykov, N.}, \bibinfo{author}{Batra, S.},
  \bibinfo{author}{Bhargava, P.}, \bibinfo{author}{Bhosale, S.}, et~al.,
  \bibinfo{year}{2023}.
\newblock \bibinfo{title}{Llama 2: Open foundation and fine-tuned chat models}.
\newblock \bibinfo{journal}{arXiv preprint arXiv:2307.09288} .
%Type = Article
\bibitem[{Wang et~al.(2023)Wang, Sun, Li, Ouyang, Wu, Zhang, Li and
  Wang}]{wang2023gpt}
\bibinfo{author}{Wang, S.}, \bibinfo{author}{Sun, X.}, \bibinfo{author}{Li,
  X.}, \bibinfo{author}{Ouyang, R.}, \bibinfo{author}{Wu, F.},
  \bibinfo{author}{Zhang, T.}, \bibinfo{author}{Li, J.}, \bibinfo{author}{Wang,
  G.}, \bibinfo{year}{2023}.
\newblock \bibinfo{title}{Gpt-ner: Named entity recognition via large language
  models}.
\newblock \bibinfo{journal}{arXiv preprint arXiv:2304.10428} .
%Type = Article
\bibitem[{Yadav and Bethard(2019)}]{yadav2019survey}
\bibinfo{author}{Yadav, V.}, \bibinfo{author}{Bethard, S.},
  \bibinfo{year}{2019}.
\newblock \bibinfo{title}{A survey on recent advances in named entity
  recognition from deep learning models}.
\newblock \bibinfo{journal}{arXiv preprint arXiv:1910.11470} .
%Type = Article
\bibitem[{Zhang et~al.(2020a)Zhang, Qian, Wang, Zhi, Dong, Xie, Zhu and
  Jiang}]{zhang20201swasp}
\bibinfo{author}{Zhang, B.}, \bibinfo{author}{Qian, S.B.},
  \bibinfo{author}{Wang, J.J.}, \bibinfo{author}{Zhi, Q.J.},
  \bibinfo{author}{Dong, A.J.}, \bibinfo{author}{Xie, W.},
  \bibinfo{author}{Zhu, L.Y.}, \bibinfo{author}{Jiang, L.Q.},
  \bibinfo{year}{2020}a.
\newblock \bibinfo{title}{1swasp j034439. 97+ 030425.5: a short-period
  eclipsing binary system with a close-in stellar companion}.
\newblock \bibinfo{journal}{Research in Astronomy and Astrophysics}
  \bibinfo{volume}{20}, \bibinfo{pages}{047}.
%Type = Article
\bibitem[{Zhang et~al.(2020b)Zhang, Chen, Huo, Zhang, Xiang, Yuan, Huang, Wang
  and Liu}]{zhang2020catalogue}
\bibinfo{author}{Zhang, M.}, \bibinfo{author}{Chen, B.Q.},
  \bibinfo{author}{Huo, Z.Y.}, \bibinfo{author}{Zhang, H.W.},
  \bibinfo{author}{Xiang, M.S.}, \bibinfo{author}{Yuan, H.B.},
  \bibinfo{author}{Huang, Y.}, \bibinfo{author}{Wang, C.},
  \bibinfo{author}{Liu, X.W.}, \bibinfo{year}{2020}b.
\newblock \bibinfo{title}{A catalogue of h$\alpha$ emission-line point sources
  in the vicinity fields of m 31 and m 33 from the lamost survey}.
\newblock \bibinfo{journal}{Research in Astronomy and Astrophysics}
  \bibinfo{volume}{20}, \bibinfo{pages}{097}.
%Type = Article
\bibitem[{ZHAO et~al.(2023)ZHAO, LU, DENG, ZHENG, WANG, CHOWDHURY, YUN, CUI,
  XUCHAO, ZHAO et~al.}]{zhao2023domain}
\bibinfo{author}{ZHAO, X.}, \bibinfo{author}{LU, J.}, \bibinfo{author}{DENG,
  C.}, \bibinfo{author}{ZHENG, C.}, \bibinfo{author}{WANG, J.},
  \bibinfo{author}{CHOWDHURY, T.}, \bibinfo{author}{YUN, L.},
  \bibinfo{author}{CUI, H.}, \bibinfo{author}{XUCHAO, Z.},
  \bibinfo{author}{ZHAO, T.}, et~al., \bibinfo{year}{2023}.
\newblock \bibinfo{title}{Domain specialization as the key to make large
  language models disruptive: A comprehensive survey}.
\newblock \bibinfo{journal}{arXiv preprint arXiv:2305.18703} .
%Type = Inproceedings
\bibitem[{Zhao et~al.(2021)Zhao, Wallace, Feng, Klein and
  Singh}]{zhao2021calibrate}
\bibinfo{author}{Zhao, Z.}, \bibinfo{author}{Wallace, E.},
  \bibinfo{author}{Feng, S.}, \bibinfo{author}{Klein, D.},
  \bibinfo{author}{Singh, S.}, \bibinfo{year}{2021}.
\newblock \bibinfo{title}{Calibrate before use: Improving few-shot performance
  of language models}, in: \bibinfo{booktitle}{International Conference on
  Machine Learning}, \bibinfo{organization}{PMLR}. pp.
  \bibinfo{pages}{12697--12706}.

\end{thebibliography}

%% else use the following coding to input the bibitems directly in the
%% TeX file.

%%\begin{thebibliography}{00}

%% \bibitem[Author(year)]{label}
%% For example:

%% \bibitem[Aladro et al.(2015)]{Aladro15} Aladro, R., Martín, S., Riquelme, D., et al. 2015, \aas, 579, A101

%%\end{thebibliography}

\end{document}